\documentclass[sigconf]{acmart}
\renewcommand\footnotetextcopyrightpermission[1]{}
\usepackage{amsmath}
\usepackage{multirow}
\usepackage{bm}
\usepackage{bbm}
\usepackage[noend]{algpseudocode}
\usepackage{algorithmicx,algorithm}
\usepackage{subfigure}

\AtBeginDocument{%
  \providecommand\BibTeX{{%
    \normalfont B\kern-0.5em{\scshape i\kern-0.25em b}\kern-0.8em\TeX}}}

\setcopyright{acmcopyright}
\copyrightyear{2022}
\acmYear{2022}
\acmDOI{10.1145/1122445.1122456}

\acmConference[2022]{Unpublished work}{June, 2022}{China}
\acmBooktitle{Unpublished work, 2022}
\acmPrice{15.00}
\acmISBN{978-1-4503-XXXX-X/18/06}

\title{Contrastive Information Transfer for Pre-Ranking Systems}

\author{Yue Cao\footnotemark[1], XiaoJiang Zhou\footnotemark[1], Peihao Huang, Yao Xiao, Dayao Chen, Sheng Chen}

\affiliation{%
  \institution{Meituan Inc.}
  \city{Beijing}
  \country{P.R. China}}
\email{yuecao@pku.edu.cn, {zhouxiaojiang, huangpeihao, xiaoyao06, chendayao, chensheng19}@meituan.com}

\begin{document}

%%
%% By default, the full list of authors will be used in the page
%% headers. Often, this list is too long, and will overlap
%% other information printed in the page headers. This command allows
%% the author to define a more concise list
%% of authors' names for this purpose.
\renewcommand{\shortauthors}{Cao et al.}

%%
%% The abstract is a short summary of the work to be presented in the
%% article.
\begin{abstract}
Real-word search and recommender systems usually adopt a multi-stage ranking architecture, including matching, pre-ranking, ranking, and re-ranking. Previous works mainly focus on the ranking stage while very few focus on the pre-ranking stage. In this paper, we focus on the information transfer from ranking to pre-ranking stage. We propose a new Contrastive Information Transfer (CIT) framework to transfer useful information from ranking model to pre-ranking model. We train the pre-ranking model to distinguish the positive pair of representation from a set of positive and negative pairs with a contrastive objective. As a consequence, the pre-ranking model can make full use of rich information in ranking model's representations. The CIT framework also has the advantage of alleviating selection bias and improving the performance of recall metrics, which is crucial for pre-ranking models. We conduct extensive experiments including offline datasets and online A/B testing. Experimental results show that CIT achieves superior results than competitive models. In addition, a strict online A/B testing at one of the world's largest E-commercial platforms shows that the proposed model achieves 0.63\% improvements on CTR and 1.64\% improvements on VBR. The proposed model now has been deployed online and serves the main traffic of this system, contributing a remarkable business growth.
\end{abstract}

%%
%% The code below is generated by the tool at http://dl.acm.org/ccs.cfm.
%% Please copy and paste the code instead of the example below.
%%
\begin{CCSXML}
<ccs2012>
<concept>
<concept_id>10002951.10003317.10003338.10003343</concept_id>
<concept_desc>Information systems~Learning to rank</concept_desc>
<concept_significance>500</concept_significance>
</concept>
<concept>
<concept_id>10002951.10003317.10003347.10003350</concept_id>
<concept_desc>Information systems~Recommender systems</concept_desc>
<concept_significance>500</concept_significance>
</concept>
</ccs2012>
\end{CCSXML}

\ccsdesc[500]{Information systems~Learning to rank}
\ccsdesc[500]{Information systems~Recommender systems}

%%
%% Keywords. The author(s) should pick words that accurately describe
%% the work being presented. Separate the keywords with commas.
\keywords{Learning to Rank, Pre-Ranking, Contrastive Learning, Search Systems}

%% A "teaser" image appears between the author and affiliation
%% information and the body of the document, and typically spans the
%% page.
% \begin{teaserfigure}
%   \includegraphics[width=\textwidth]{sampleteaser}
%   \caption{Seattle Mariners at Spring Training, 2010.}
%   \Description{Enjoying the baseball game from the third-base
%   seats. Ichiro Suzuki preparing to bat.}
%   \label{fig:teaser}
% \end{teaserfigure}

%%
%% This command processes the author and affiliation and title
%% information and builds the first part of the formatted document.
\maketitle
\footnotetext[1]{The first two authors contributed equally to this paper.}

\section{Introduction}
\begin{figure}[tb]
	\centering 
	\includegraphics[width=0.8\linewidth]{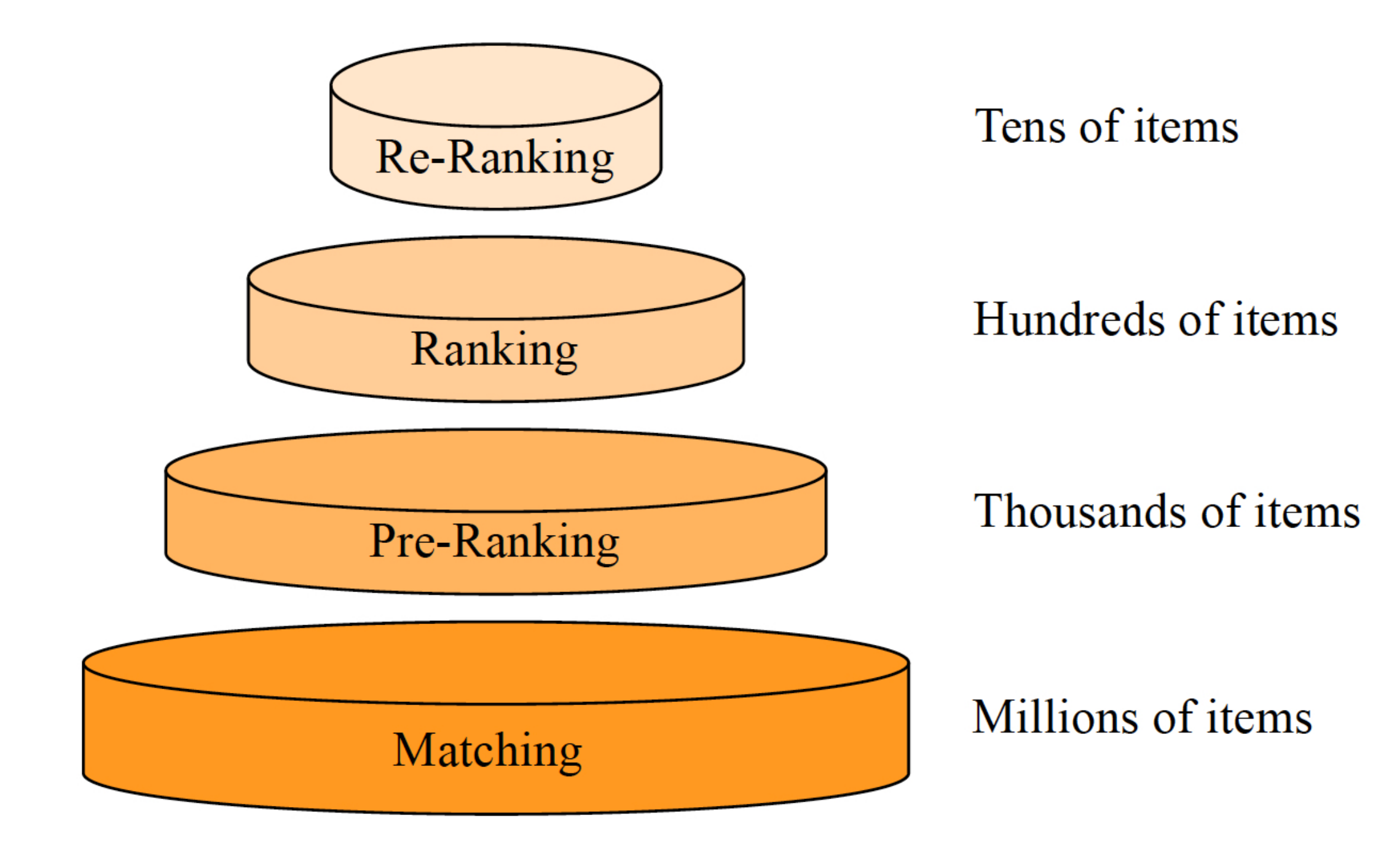}  
	\caption{Illustration of a real-world multi-stage search system.}  
	\label{fig1}   
\end{figure}

With the rapid growth of internet services, search systems are becoming increasingly important in assisting users to find what they want.
For the balance of performance and efficiency, industrial search systems usually consist of multiple cascade stages \cite{DBLP:journals/corr/abs-2007-16122}: matching, pre-ranking, ranking, and re-ranking, as shown in Fig~\ref{fig1}. Previous work mainly pay attention to the ranking stage \cite{DBLP:conf/recsys/CovingtonAS16, DBLP:journals/corr/GaiZLLW17, DBLP:conf/kdd/GrbovicC18, DBLP:conf/aaai/LyuDHR20, DBLP:journals/corr/ZhouSZMYDZJLG17, DBLP:conf/aaai/ZhouMFPBZZG19}, while very few focus on the pre-ranking stage.
This paper focuses on the \textbf{pre-ranking} stage, where the model receives thousands of candidate items retrieved from the matching stage, and applies a simple model to rank and select hundreds of candidates for the subsequent ranking model.

To meet the efficiency constraint, the pre-ranking model is usually a simpler model compared with the ranking model. For a long time, the pre-ranking model is usually designed like the tree model or shallower linear models \cite{DBLP:conf/kdd/McMahanHSYEGNPDGCLWHBK13, DBLP:journals/corr/abs-2007-16122}. 
% With the development of deep learning, deep two-tower models \cite{DBLP:conf/recsys/CovingtonAS16} are proposed. The two-tower models use two separate deep neural networks to encode context features (including request, user features, etc.) and item features, and compute the inner product of the two representations as the pre-ranking score. Although achieving promising results, two-tower models lack the interaction between user and item representations, resulting in inferior results \cite{DBLP:conf/kdd/ZhuLZLHLG18}. To model the interactions, COLD \cite{DBLP:journals/corr/abs-2007-16122} merges the two towers into a single tower, and feeds user and item features together into a unified deep neural model. How to learn better representations is the key point for pre-ranking models.
Recently, some works focus on leveraging knowledge from the ranking model to assist the pre-ranking model \cite{DBLP:conf/kdd/XuLGGYPSWSO20}.

In this paper, we reveal that the \textbf{ranking consistency between pre-ranking and ranking stages is a key factor to the whole system's performance.} From the perspective of the whole system, the top-$u$ items retrieved by the pre-ranking model will be re-ranked by the subsequent ranking model and therefore, the relative orders within top-$u$ items predicted by the pre-ranking model will not affect the final result.

To this end, we propose a novel \textbf{Contrastive Information Transfer (CIT)} framework to improve the ranking consistency between pre-ranking and ranking models. To optimize the CIT objective, we design labels that takes the orders of results of the ranking model into account, which enables the pre-ranking model to focus more on top items, thereby improving the performance on recall metric. Then, we train the pre-ranking model to distinguish the positive pair from a set of positive and negative pairs with a contrastive objective. Theoretically, CIT maximizes a lower bound of mutual information between pre-ranking model's representations and ranking model's representation, thus facilitating the pre-ranking to make full use of rich information in representations and orders of the ranking model.

% Denote the pre-ranking model as $\phi$ and ranking model as $\phi^T$. For a positive example $\{(x_0, y_0)|y_0=1\}$ and $K$ negative examples $\{(x_1, y_1), (x_2, y_2),$ $\cdots, (x_K, y_K)|y_k=0\}$, we treat $(\phi(q, x_0), \phi^T(q, x_0))$ as a positive pair, and 
% $\{(\phi(q, x_0), \phi(q, x_j))_{j=1}^{K}\}$ as negative pairs, and train the pre-ranking model to distinguish the positive pair from a set of positive and negative pairs with a contrastive objective. Theoretically, CIT maximizes a lower bound of mutual information between pre-ranking model's representations and ranking model's representation, thus facilitating the pre-ranking to make full use of rich information in representations of the ranking model.
% We also propose a new way to design labels that takes the orders of results of the ranking model into account, which enables the pre-ranking model to focus more on top items, thereby improving the performance on recall metric.
% Compared with knowledge transfer frameworks such as knowledge distillation \cite{DBLP:journals/corr/HintonVD15}, CIT is better at transferring structured knowledge, and can significantly improve the recall rate.

We conduct extensive experiments in one of the world's largest E-commercial platforms. CIT achieves 0.63\% improvement in Click-Through Rate (CTR) and 1.64\% improvement in Visit-Buy Rate (VBR), which is very significant considering the huge turnover of this platforms. \textit{Since 2022, CIT has been deployed online and served the main search traffic of this system, and bringing significant online profit improvement.}

In sum, the prime contributions of this paper can be summarized as follows:
\begin{itemize}
\item We propose a novel contrastive information transfer framework that transfers rich information of representations from ranking model to pre-ranking model.
\item Extensive experiments including offline dataset and online A/B test show the superior of CIT compared with competitive methods. Moreover, CIT has been successfully deployed online and serves the main search traffic.
\end{itemize}

\begin{figure*}[tb]
	\centering 
	\includegraphics[width=0.8\linewidth]{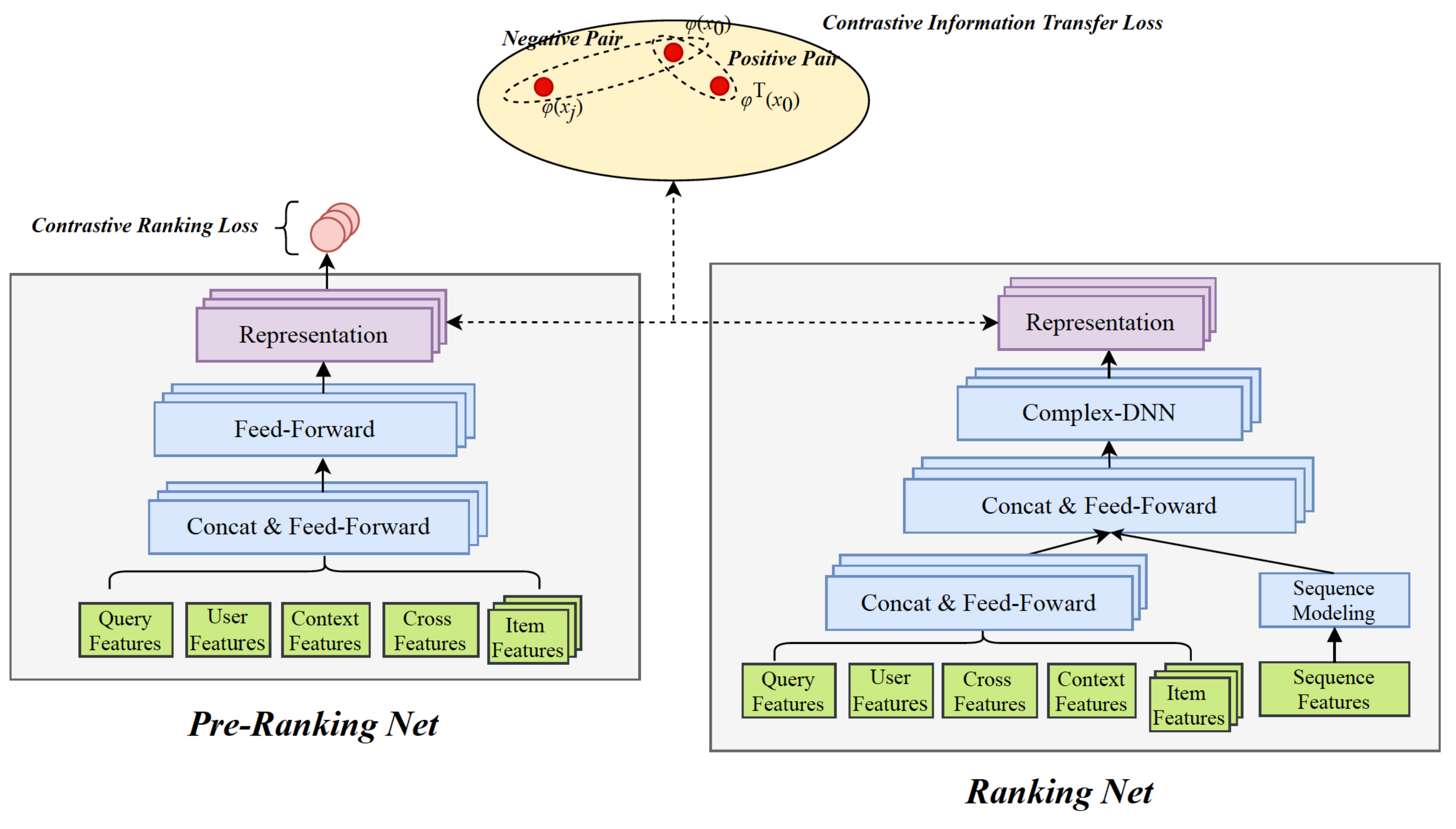}  
	\caption{Overall architecture of the proposed CIT framework. The left part is the pre-ranking net and the right part is the ranking net. During training, the gradients will be not passed to the ranking model and the parameters of the ranking model are fixed.}  
	\label{fig2}   
\end{figure*}

\section{Methodology}
\subsection{Overview}
% As mentioned above, A typical industrial system consists of four cascade stages: matching, pre-ranking, ranking, and re-ranking. 
A high-level overview of the pre-ranking and ranking models in our system are shown in Figure.~\ref{fig2}
The pre-ranking model in our system consists of an embedding and several feed-forward layers. Our ranking model is also built upon deep neural networks, while the architecture of the ranking model is more complicated than that of the pre-ranking model for higher accuracy. 
% The embedding layer is used to transform one-hot features into low dimensional dense representations, and feed-forward layers are used to automatically learn feature interactions and predict the probability of a sample $x$ being clicked. 

% The pairwise loss is used to optimize the pre-ranking model\footnote{We have also tried to use point-wise loss, but it performs worse than pair-wise loss in Eq.~\ref{obj1}.}.
% Suppose $(x_1, x_2, \cdots, x_n)$ is the list of input features under the same request, and $(y_1, y_2, \cdots, y_n)$ is the corresponding labels. We denote $p_{ij}$ as the probability of $x_i$ being ranked before $x_j$, which is calculated as:
% \begin{equation}
%     p_{ij} = \frac{1}{1+e^{-\sigma (s_i - s_j)}}
% \end{equation}
% where $\sigma$ is a hyper-parameter, and $s_i, s_j$ are outputs of the network after the sigmoid activation.

% We denote $\hat{p}_{ij}$ as the gold label of $x_i$ being ranked before $x_j$, i.e., $\hat{p}_{ij}=\mathbbm{1}_{y_i > y_j}$. The training objective is to minimize the following cross-entropy loss:
% \begin{equation}
%     \mathcal{L} = -\sum_{i=1}^{n}\sum_{j=1}^{n} \left[\hat{p}_{ij}\log p_{ij} + (1-\hat{p}_{ij})\log(1-p_{ij}) \right] \label{obj1}
% \end{equation}
   
Our ranking model is also built upon deep neural networks, while the architecture of the ranking model is more complicated than that of the pre-ranking model for higher accuracy. 
% In addition to the embedding layer and feed-forward layers, our ranking model includes a sequence modeling layer \cite{DBLP:conf/kdd/ZhouZSFZMYJLG18} for extracting users' historical interests, as well as an MMoE layer \cite{DBLP:conf/recsys/TangLZG20, DBLP:conf/kdd/MaZYCHC18} for multi-business learning. 
% The ranking model is optimized with the pairwise loss.
In our system, the pre-ranking model ranks 2,000 candidates retrieved from the matching model, and selects top-150 candidates for the ranking model.

\subsection{Learning from Ranking Model via Contrastive Information Transfer}
Compared with the ranking model, the pre-ranking model has similar training data, training objectives, and architectures. The difference is that the ranking model is more sophisticated in architecture and contains more parameters. Therefore, it is natural to transfer the information from the ranking model to the pre-ranking model.
For this purpose, we propose a novel \textbf{Contrastive Information Transfer (CIT)} framework in this paper.

As shown in Fig~\ref{fig2}, we introduce the ranking model into the computation graph of the pre-ranking model.
We regard representations of the same positive example encoded by pre-ranking and ranking models as positive pairs, and representations of a positive and negative example encoded by pre-ranking model respectively as negative pairs. 
% By learning to distinguish positive pairs from a set of negative pairs, the pre-ranking model is facilitated to exploit the useful information in ranking model's representations.
Notice that the ranking model is only used to provide representations, and its parameters will not be updated during training. 
At the inference phase, we only export the pre-ranking part and deploy it online.

\subsubsection{Formulation}
Denote $\phi^T$ as the ranking model and $\phi$ as the pre-ranking model. Given a positive instance $\{(x_0, y_0)|y_0=1\}$, we feed $x_0$ into the pre-ranking and ranking model respectively, and obtain the corresponding representations $\phi(q, x_0)$ and $\phi^T(q, x_0)$. At the same time, we sample $K$ negative instances from the same request, and feed these instances into the pre-ranking model to obtain the corresponding representations $\{\phi(q, x_1), \phi(q, x_2), \cdots, \phi(q, x_K)$ $|y_{1,2,\cdots,K}=0\}$.

We treat $(\phi(q, x_0),\phi^T(q, x_0))$ as the positive pair, and $\{(\phi(q, x_0),$ \\
$\phi(q, x_j))_{j=1}^{K}\}$ as negative pairs, and minimize the following contrastive loss:
\begin{equation}
    \mathcal{L}_{CKT} = -\sum_{q \in Q} \left[ \log \frac{\exp{\left(\langle \phi(q, x_0),\phi^T(q, x_0)\rangle / \tau^\prime\right)}}{\sum_{j=0}^{K}\exp{\left(\langle \phi(q, x_0),\phi(q, x_j)\rangle / \tau^\prime\right)}} \right] \label{ckt}
\end{equation}
where $\langle \cdot,\cdot \rangle$ refers to the inner product, and $\tau^\prime$ is the temperature. After incorporating the CIT loss, the total training objective of our pre-ranking model can be formulated as:
\begin{equation}
    \mathcal{L} = \lambda \mathcal{L}_{CE} + (1-\lambda) \mathcal{L}_{CKT} \label{obj2}
\end{equation}
where $\mathcal{L}_{CE}$ is the cross-entropy loss, and $\lambda$ is the hyper-parameter to balance two loss terms.

Intuitively, Eq.~\ref{ckt} encourages the pre-ranking model to find the most similar one from a set of representations, which is defined as the corresponding ranking model's representation. Therefore the pre-ranking model is facilitated to exploit the information from ranking model.   
Theoretically, minimizing Obj.~\ref{ckt} can also be interpreted as maximizing the lower bound of mutual information between pre-ranking model's representation $\phi(x_0)$ and ranking model's representation $\phi^T(x_0)$ \cite{DBLP:journals/corr/abs-1807-03748}:
\begin{equation}
    \begin{split}
        \mathcal{L}_{CKT} &= -\mathbb{E}_{Q} \left[ \log \frac{\exp{\left(\langle \phi(q, x_0),\phi^T(q, x_0)\rangle / \tau^\prime\right)}}{\sum_{j=0}^{K}\exp{\left(\langle \phi(q, x_0),\phi(q, x_j)\rangle / \tau^\prime\right)}} \right] \\
       & \geq - I(\phi(q, x_0),\phi^T(q, x_0)) + \log(N) \label{eq2}
    \end{split}
\end{equation}
where $I(\cdot, \cdot)$ refers to the mutual information. Therefore, the proposed method maximizes the agreement of representations of pre-ranking and ranking at the mutual information level.
We named the proposed method as \textbf{Contrastive Information Transfer (CIT)}.

\subsubsection{Design labels $y$} \label{sec3.3.2}

% \begin{figure}[tb]
% 	\centering 
% 	\includegraphics[width=0.7\linewidth]{figures/Fig3.pdf}  
% 	\caption{The design of labels $y$ for contrastive information transfer.} 
% % 	The pre-ranking model rank $v$ items retrieved from the matching stage and select top-$u$ items for the ranking model. We assign $y=1$ for top-$u$ items ranked by the ranking model and $y=0$ for rest $v-u$ items. }  
% 	\label{fig3}   
% \end{figure}

The choices of label $y$ are crucial for CIT. A straightforward approach is to treat the clicked items as positive instances ($y=1$), while the displayed items but not clicked as negative instances ($y=0$). However, as the ranking model is available, this approach does not use any information of orders of results of the ranking model.

Let's look back at the goal and nature of the pre-ranking system. In a real-world search system, the pre-ranking system is used to retrieve top-$u$ items from $v (v \gg u)$ candidates for the subsequent ranking model.  
From the perspective of the whole system, the relative orders of top-$u$ items of pre-ranking do not affect the final result, because they will all be re-ranked by the subsequent ranking model. \textbf{Therefore, we should care more about the recall rate of top-$u$ items.}

% \textbf{While previous works model the pre-ranking as a click-through rate (CTR) estimator task, we argue that the pre-ranking should be modeled as a retrieval task.}
In this paper, we propose an approach which is more in line with the nature of pre-ranking: we treat top-$u$ items ranked by the ranking model as positive examples ($y=1$), and the rest $v-u$ items as negative examples ($y=0$).
% We will conduct experiments to demonstrate the effectiveness of this approach.

\subsubsection{CIT can alleviate the Selection Bias}
We argue that the proposed contrastive information transfer approach can alleviate the \textbf{selection bias} issue \cite{DBLP:conf/cikm/YuanHYZCDL19} of the pre-ranking model to some extent. 

The pre-ranking model suffers from two selection biases: (1) without considering non-displayed items, and (2) without considering the request that has no positive feedback. For pairwise objectives, requests without positive feedback can not be used for training and have to be discarded. 

The presence of selection bias makes the distribution of training data and real data inconsistent.
Previous work \cite{DBLP:conf/cikm/YuanHYZCDL19} has confirmed that ignoring these selection biases may hurt the training and reduce the performance.
CIT considered the entire 2,000 items, which may contain non-displayed items for some request, so it can alleviate the first type of selection bias.
Besides, the introduction of $\mathcal{L}_{CKT}$ enables us to utilize requests without positive feedback\footnote{For requests without positive feedback, we only minimize the second term in Eq.~\ref{obj2}.}. Therefore, CIT can alleviate the second type of selection bias, either.

\subsubsection{Advantages over Knowledge Distillation}
Knowledge distillation \cite{DBLP:journals/corr/HintonVD15} refers to transfer the dark knowledge from one network to another network.

The idea behind CKT is similar to knowledge distillation, but it has the following advantages in our scenario: (1) CIT pays more attention to top results (positive instances), which is more in line with the nature of the pre-ranking model, i.e., select top candidates for the ranking model. (2) The data of a real-world search system is highly structured, i.e., dependencies exist in different dimensions of features. While the commonly used KL divergence in knowledge distillation treats all dimensions as independent, thus is insufficient for capturing structural knowledge \cite{DBLP:conf/cvpr/00010G0HC21, DBLP:conf/iclr/TianKI20}. In contrast, CIT maximizes the lower-bound of the mutual information between two representations, which is more advantageous in dealing with higher-order correlations.

% We will also conduct experiments to demonstrate the superiority of CIT over knowledge distillation-based methods. 

\section{Experiments}
% In this section, we present the experiment details. Firstly, we introduce the datasets. Then we introduce the experimental metrics, experimental settings and implemental details. Finally, we list the competitive models compared with our proposed model.
\subsection{Experiment Setups}
\subsubsection{Dataset}
\textbf{To the best of our knowledge, there is no available benchmark for the pre-ranking task, and previous works \cite{DBLP:journals/corr/abs-2007-16122, DBLP:conf/sigir/MaWZLZLLXZ21} conduct experiments on their own datasets\footnote{Their datasets are commercial and not publicly available.}}. 
To verify the effectiveness of the proposed method, we follow previous work to conduct experiments using real-world data collected from our own system.

The dataset is collected from the searching system in one of the world's largest E-commercial platforms. This dataset contains more than 100 million users and more than 10 billion training examples. 
% We digitize each example through feature extraction pipelines and label it according to business rules. 
% To implement CIT, we label the example not only based on user feedback but also orders of the ranking model, which may contains non-displayed examples. 
% Displayed example means the example is shown to the front user. 
% In our scenario, the number of displayed examples is hundreds of times larger than non-displayed examples. 
Following the classic setting for industrial models, we treat the data of the first 14 days as the training set, the data of the following 2 days as validation and test set respectively.

\subsubsection{Evaluation Metrics}
Following previous works, we use \textbf{G-AUC (Group-Area Under Curve)} \cite{DBLP:conf/kdd/ZhuJTPZLG17}, \textbf{Recall} \cite{DBLP:journals/corr/abs-2007-16122}, and \textbf{NDCG@k (Normalized Discounted Cumulative Gain)} for offline evaluations. We use $k=15$ because NDCG@15 is the core metric in our business.
% Since the results of pre-ranking will be re-ranked by the subsequent ranking model, the improvement of NDCG and AUC metrics may not affect the final result from the perspective of the whole. For example, reversing the order of the items ranked first and second by the pre-ranking model, both NDCG and AUC will change significantly. But as they are both selected for the next stage and will be re-ranked by the ranking model, the final orders of the whole search system will not change. \textbf{Therefore, for pre-ranking systems, we care more about the Recall metric}.
For online experiments, we use \textbf{CTR (Click-Through Rate)} and \textbf{VBR (Visited-Buy Rate)} as online metrics.
% CTR measures the click rate of users, which is calculated as the ratio of users who click on a specific item to the number of total users who view an item, and VBR measures the pay rate of uses.
% , which is calculated as the ratio of users who buy a specific item to the number of total users who view an item.

% \subsection{Experimental Settings}

% \subsection{Implementation Details}
% We implement our method with the Tensorflow platform\footnote{\url{https://github.com/tensorflow/tensorflow}}. 

% For training, we use the Adagrad optimizer \cite{DBLP:conf/colt/DuchiHS10}. We use the exponential-decay learning rate schedule \cite{DBLP:conf/iclr/0005A20} with a learning rate of 0.02. The batch size for training is set to 64. We train the model on a CPU machine with 256GB RAM, while it usually takes about 20 hours for the model to converge. According to the performance, we set $\tau=0.1$, $\tau^\prime=0.1$, and $\lambda=0.7$.  

% We constructed hundreds of features for the pre-ranking model. For discrete features, we convert them into one-hot vectors. As for id features, we hash it firstly, then process like discrete features. And for continuous features, we first bucket it to discrete features. Then all one-hot vectors adopt an embedding layer to convert them into dense vectors separately. An aggregate layer will catenate them to a big embedding. A three-layer feed-forward layer is adopted to model feature interactions.
% Finally, 3 fc(full connect) layers add to the model, which makes the model bigger capacity. In our setting, the hidden size for fc layers is set to 512, and the embedding size is set to 16.

\subsection{Competitive Models}
Following previous work \cite{DBLP:journals/corr/abs-2007-16122, DBLP:conf/sigir/MaWZLZLLXZ21}, we compare CIT with following popular pre-ranking models.
% Additionally, to verify the effectiveness of each module of CIT, we also test some variations of CIT.
\begin{itemize}
    \item Classic pre-ranking models, including \textbf{Vector-Product based Deep Models (VPDM)} \cite{DBLP:conf/www/YangYCHLWXC20, DBLP:journals/corr/abs-2007-16122}, \textbf{Deep Neural Networks (DNN)}, and \textbf{Deep Neural Networks with Knowledge Distillation (DNN+KD)}. DNN+KD is the previous pre-ranking model in our system.
    
    \item Famous pre-ranking model proposed by others, including \textbf{COLD} \cite{DBLP:journals/corr/abs-2007-16122}, and \textbf{FSCD} \cite{DBLP:conf/sigir/MaWZLZLLXZ21}\footnote{We implement COLD and FSCD upon the DNN architecture.}. 
    % For a detailed introduction of these models, please refer to Related Work.
    Please Note that COLD and FSCD mainly focus on how to select features for pre-ranking models, which is different from us. Therefore, it is less necessary to compare CIT with them. But since they are classic models in the pre-ranking field, we also list their results as a reference.
    
    % \item 
    % \textbf{CIT w/o CR}: remove the contrastive ranking loss in CIT. 
    % \textbf{CIT (point-wise)}: replace the contrastive ranking loss in CIT with point-wise loss, which is implemented as the cross-entropy loss. \textbf{CIT (pair-wise)}: replace the contrastive ranking loss in CIT with traditional pair-wise loss in Eq.~\ref{logpij}. 
    
    \item \textbf{CIT w/o CIT}: 
    to demonstrate the superiority of CIT over knowledge distillation, we set a variation that replaces the CIT module with traditional knowledge distillation, where the model optimizes the following KD loss \cite{DBLP:journals/corr/HintonVD15}:
    \begin{equation}
        \mathcal{L}_{KD} = \mathbf{KL}(\phi^T(q,x)||\phi(q,x))
    \end{equation}
    For a fair comparison, we distill all top-$u$ examples, therefore the training data of KD is the same with CIT.
    
    To illustrate that the improvement is not entirely attributable to the choice of label $y$, we also set a variation that replaces the CIT with a pairwise-based loss where the choice of label is the same as in Section~\ref{sec3.3.2}, i.e., $y=1$ for top-$u$ items and $y=0$ for the rest.
\end{itemize}

\textbf{The above experimental settings, including competitive models and datasets chosen, are completely consistent with previous work} \cite{DBLP:conf/sigir/MaWZLZLLXZ21, DBLP:journals/corr/abs-2007-16122, DBLP:conf/kdd/XuLGGYPSWSO20}.

\section{Results and Analysis}
\subsection{Overall Results}
\begin{table}[t!]
\small
\begin{center}
\setlength{\tabcolsep}{6.0mm}{
\begin{tabular}{l|ccc}
\toprule \bf Method & \bf NDCG & \bf G-AUC & \bf Recall \\ \midrule
% {LR} & 0.6938 & 0.8575 & 0.5942 \\ 
{VPDM} & 0.6988* & 0.8633* & 0.5948* \\ 
{DNN} & 0.7123* & 0.8772* & 0.6321* \\
{DNN+KD} & 0.7134* & 0.8846* & 0.6359* \\ \hline
{COLD} &0.7125* & 0.8777* & 0.6325* \\ 
{FSCD} &0.7129* & 0.8780* & 0.6331*  \\ \hline
{CIT} & \bf 0.7222 & \bf 0.8907 & \bf 0.7481 \\
\bottomrule
\end{tabular}}
\end{center}
\caption{\label{result:1} Performance comparison of different model on offline dataset.
"*" indicates that the improvement of CIT over this baseline is statistically significant at p-value < 0.05 over paired Wilcoxon-test.}
\end{table}
The overall results of different models on offline dataset are shown in Table~\ref{result:1}. In sum, CIT outperforms competitive models in NDCG, G-AUC, and Recall metrics concurrently. Particularly, the improvement of CIT on Recall is very significant, with an increase of more than 0.1 absolute scores (17.6\%) compared to baseline models.

Compared with the previous online pre-ranking model DNN+KD, CIT achieves 1.23\% improvement on NDCG (0.7222 vs. 0.7134), 0.69\% improvement on G-AUC (0.8907 vs. 0.8846), and 17.6\% on Recall (0.7481 vs. 0.6359). 

Compared with DNN, neither COLD nor FSCD achieves remarkable improvement. This is because COLD and FSCD focus on selecting important features for pre-ranking models, while in our scenario, the importance of features can mostly be derived from prior knowledge\footnote{The initial feature set of COLD and FSCD is the same with CIT and DNN.}. Besides, we found in experiments that COLD and FSCD may miss some intuitively important features.
CIT essentially changes the way of modeling and training pre-ranking models, making the optimization of the pre-ranking model more in line with the goal and nature of pre-ranking. Therefore, CIT can achieve significant improvement over all competitive models.

\subsection{Ablation Study of CIT}
% \begin{table}[t!]
% \begin{center}
% \setlength{\tabcolsep}{2.5mm}{
% \begin{tabular}{l|ccc}
% \toprule \bf Method & \bf NDCG & \bf G-AUC & \bf Recall \\ \midrule
% {CIT} & 0.7222 & 0.8907 & \bf 0.7481 \\ 
% {CIT (point-wise)} & 0.7208 & \bf 0.8996 & 0.7171 \\ 
% {CIT (pair-wise)} & 0.7268 & 0.8893 & 0.7298 \\ 
% {CIT w/o CIT} & - & - & - \\ 
% {CIT (KD)} & 0.7261 & 0.8830 & - \\ 

% \bottomrule
% \end{tabular}}
% \end{center}
% \caption{\label{result:2} Ablation study of CIT.}
% \end{table}

	\begin{table}[t!]
	    \small
		\begin{center}
			\setlength{\tabcolsep}{2.2mm}{
				\begin{tabular}{c|c|ccc}
					\toprule
					\multicolumn{2}{c|}{\bf Method} & \bf NDCG & \bf G-AUC & \bf Recall \\  \midrule
                
                    \multicolumn{2}{c|}{\bf CIT} & \bf 0.7222 & 0.8907 & \bf 0.7481 \\ \hline
    %                 \multirow{3}*{\bf CIT w/o CR} & - & 0.6861 & 0.8488 & 0.7072 \\ 
				% 	~ & {+ pointwise } & 0.7092 & \bf 0.8949 & 0.7122 \\ 
				% 	~ & {+ pairwise } & 0.7204 & 0.8859 & 0.7283 \\ \hline
					\multirow{2}*{\bf CIT w/o CIT} & - & 0.7107 & 0.8704 & 0.6604 \\ 
					~ & + KD & 0.7185 & 0.8842 & 0.6675 \\ 
					~ & + pairwise & 0.7118 & 0.8687 & 0.6885 \\ 
					\bottomrule
			\end{tabular}}
		\end{center}
		\caption{\label{result:2} Ablation study of CIT.
% 		Both contrastive ranking and contrastive information transfer modules contributes to the final improvement of CIT.
		}
	\end{table}

The results of the ablation study are listed in Table~\ref{result:2}. 
% First, using contrastive ranking (CR) loss can improve NDCG and Recall scores, and the improvement in Recall metric is more significant. 
% The adoption of pointwise loss is helpful for improving G-AUC metric, but not beneficial for improving NDCG and recall metrics\footnote{We once conducted online experiments, and the model trained with pointwise loss performs worse than pairwise loss.}.

The contrastive information transfer (CIT) plays a significant role in improving the recall score, as the performance greatly decreases when the CIT module is removed. We believe that this is because $\mathcal{L}_{CIT}$ facilitates the model to learn better representations, and focuses more on top-$u$ items. The experimental results also show that the performance decrease when replace CIT with pairwise loss, indicating that the improvement is not entirely attributable to the choice of label.

\subsection{Online A/B test}
% \begin{table}[t!]
% \begin{center}
% \setlength{\tabcolsep}{3.5mm}{
% \begin{tabular}{l|cccc}
% \toprule \bf Method & \bf Recall & \bf CTR & \bf VBR \\  \midrule
% {Baseline model}  & 0.6037 & - & - \\
% {CIT} & \bf 0.7189 & \bf +0.63\% & \bf +1.64\% \\ \bottomrule
% \end{tabular}}
% \end{center}
% \caption{\label{result:ab} Online A/B testing results. Baseline model is DNN+KD, which is the last online pre-ranking model in our search system.}
% \end{table}

A strict online A/B testing is also conducted to verify the effectiveness of CIT. For online A/B testing, the baseline model is the previous online pre-ranking model in our system, which is the DNN+KD model mentioned earlier. The test model is CIT. 
The testing lasts for 14 days, with 10\% of traffic is distributed for each model respectively. 
% Note that since the results of pre-ranking will be re-ranked by the subsequent ranking and re-ranking model, the pre-ranking model usually has a smaller impact on online performance. The experimental results are shown in Table~\ref{result:ab}.

Overall, \textbf{CIT achieves 0.63\% improvement in CTR ($p$-value<0.005, confidence>0.995), and 1.64\% improvement in VBR ($p$-value<0.001, confidence>0.998)}, which can greatly increase online profit considering the large traffic of our system. Particularly, CIT greatly improves Recall@150 by an absolute 0.1152 score, from 0.6037 to 0.7189, which again verifies the effectiveness of CIT in improving results of top items.  
Besides, we find that CIT does not increase the response time (RT) of the system compared with the baseline model. This is because CIT does not introduce any additional modules or new features, but improves the training frameworks.

The results of online A/B testing demonstrate that CIT is superior to the last online pre-ranking model (DNN+KD). 
Since 2022, CIT has been deployed online and served the main traffic of our system.

\section{Conclusions}
In this paper, we propose a new contrastive information transfer framework for pre-ranking systems. The system seeks for a high recall score and is more in-line with the position of pre-ranking models. We also show that the proposed CIT framework is better for transferring structural knowledge, alleviating the selection bias, and improving the recall rate of the model. 

Future work includes employing the contrastive learning framework in other stages of the search system, such as matching, ranking, re-ranking, and so on.

%%
%% The acknowledgments section is defined using the "acks" environment
%% (and NOT an unnumbered section). This ensures the proper
%% identification of the section in the article metadata, and the
%% consistent spelling of the heading.
% \begin{acks}
% To Robert, for the bagels and explaining CMYK and color spaces.
% \end{acks}

%%
%% The next two lines define the bibliography style to be used, and
%% the bibliography file.
\bibliographystyle{ACM-Reference-Format}
\bibliography{ref}

%%% -*-BibTeX-*-
%%% Do NOT edit. File created by BibTeX with style
%%% ACM-Reference-Format-Journals [18-Jan-2012].

\begin{thebibliography}{17}

%%% ====================================================================
%%% NOTE TO THE USER: you can override these defaults by providing
%%% customized versions of any of these macros before the \bibliography
%%% command.  Each of them MUST provide its own final punctuation,
%%% except for \shownote{}, \showDOI{}, and \showURL{}.  The latter two
%%% do not use final punctuation, in order to avoid confusing it with
%%% the Web address.
%%%
%%% To suppress output of a particular field, define its macro to expand
%%% to an empty string, or better, \unskip, like this:
%%%
%%% \newcommand{\showDOI}[1]{\unskip}   % LaTeX syntax
%%%
%%% \def \showDOI #1{\unskip}           % plain TeX syntax
%%%
%%% ====================================================================

\ifx \showCODEN    \undefined \def \showCODEN     #1{\unskip}     \fi
\ifx \showDOI      \undefined \def \showDOI       #1{#1}\fi
\ifx \showISBNx    \undefined \def \showISBNx     #1{\unskip}     \fi
\ifx \showISBNxiii \undefined \def \showISBNxiii  #1{\unskip}     \fi
\ifx \showISSN     \undefined \def \showISSN      #1{\unskip}     \fi
\ifx \showLCCN     \undefined \def \showLCCN      #1{\unskip}     \fi
\ifx \shownote     \undefined \def \shownote      #1{#1}          \fi
\ifx \showarticletitle \undefined \def \showarticletitle #1{#1}   \fi
\ifx \showURL      \undefined \def \showURL       {\relax}        \fi
% The following commands are used for tagged output and should be
% invisible to TeX
\providecommand\bibfield[2]{#2}
\providecommand\bibinfo[2]{#2}
\providecommand\natexlab[1]{#1}
\providecommand\showeprint[2][]{arXiv:#2}

\bibitem[\protect\citeauthoryear{Chen, Wang, Gan, Liu, Henao, and Carin}{Chen
  et~al\mbox{.}}{2021}]%
        {DBLP:conf/cvpr/00010G0HC21}
\bibfield{author}{\bibinfo{person}{Liqun Chen}, \bibinfo{person}{Dong Wang},
  \bibinfo{person}{Zhe Gan}, \bibinfo{person}{Jingjing Liu},
  \bibinfo{person}{Ricardo Henao}, {and} \bibinfo{person}{Lawrence Carin}.}
  \bibinfo{year}{2021}\natexlab{}.
\newblock \showarticletitle{Wasserstein Contrastive Representation
  Distillation}. In \bibinfo{booktitle}{\emph{{CVPR} 2021, virtual, June 19-25,
  2021}}. \bibinfo{publisher}{Computer Vision Foundation / {IEEE}},
  \bibinfo{pages}{16296--16305}.
\newblock
\urldef\tempurl%
\url{https://openaccess.thecvf.com/content/CVPR2021/html/Chen\_Wasserstein\_Contrastive\_Representation\_Distillation\_CVPR\_2021\_paper.html}
\showURL{%
\tempurl}


\bibitem[\protect\citeauthoryear{Covington, Adams, and Sargin}{Covington
  et~al\mbox{.}}{2016}]%
        {DBLP:conf/recsys/CovingtonAS16}
\bibfield{author}{\bibinfo{person}{Paul Covington}, \bibinfo{person}{Jay
  Adams}, {and} \bibinfo{person}{Emre Sargin}.}
  \bibinfo{year}{2016}\natexlab{}.
\newblock \showarticletitle{Deep Neural Networks for YouTube Recommendations}.
  In \bibinfo{booktitle}{\emph{Proceedings of the 10th {ACM} Conference on
  Recommender Systems, Boston, MA, USA, September 15-19, 2016}}.
  \bibinfo{publisher}{{ACM}}, \bibinfo{pages}{191--198}.
\newblock
\urldef\tempurl%
\url{https://doi.org/10.1145/2959100.2959190}
\showDOI{\tempurl}


\bibitem[\protect\citeauthoryear{Gai, Zhu, Li, Liu, and Wang}{Gai
  et~al\mbox{.}}{2017}]%
        {DBLP:journals/corr/GaiZLLW17}
\bibfield{author}{\bibinfo{person}{Kun Gai}, \bibinfo{person}{Xiaoqiang Zhu},
  \bibinfo{person}{Han Li}, \bibinfo{person}{Kai Liu}, {and}
  \bibinfo{person}{Zhe Wang}.} \bibinfo{year}{2017}\natexlab{}.
\newblock \showarticletitle{Learning Piece-wise Linear Models from Large Scale
  Data for Ad Click Prediction}.
\newblock \bibinfo{journal}{\emph{CoRR}}  \bibinfo{volume}{abs/1704.05194}
  (\bibinfo{year}{2017}).
\newblock
\showeprint[arXiv]{1704.05194}
\urldef\tempurl%
\url{http://arxiv.org/abs/1704.05194}
\showURL{%
\tempurl}


\bibitem[\protect\citeauthoryear{Grbovic and Cheng}{Grbovic and Cheng}{2018}]%
        {DBLP:conf/kdd/GrbovicC18}
\bibfield{author}{\bibinfo{person}{Mihajlo Grbovic} {and}
  \bibinfo{person}{Haibin Cheng}.} \bibinfo{year}{2018}\natexlab{}.
\newblock \showarticletitle{Real-time Personalization using Embeddings for
  Search Ranking at Airbnb}. In \bibinfo{booktitle}{\emph{{KDD} 2018, London,
  UK, August 19-23, 2018}}, \bibfield{editor}{\bibinfo{person}{Yike Guo} {and}
  \bibinfo{person}{Faisal Farooq}} (Eds.). \bibinfo{publisher}{{ACM}},
  \bibinfo{pages}{311--320}.
\newblock
\urldef\tempurl%
\url{https://doi.org/10.1145/3219819.3219885}
\showDOI{\tempurl}


\bibitem[\protect\citeauthoryear{Hinton, Vinyals, and Dean}{Hinton
  et~al\mbox{.}}{2015}]%
        {DBLP:journals/corr/HintonVD15}
\bibfield{author}{\bibinfo{person}{Geoffrey~E. Hinton}, \bibinfo{person}{Oriol
  Vinyals}, {and} \bibinfo{person}{Jeffrey Dean}.}
  \bibinfo{year}{2015}\natexlab{}.
\newblock \showarticletitle{Distilling the Knowledge in a Neural Network}.
\newblock \bibinfo{journal}{\emph{CoRR}}  \bibinfo{volume}{abs/1503.02531}
  (\bibinfo{year}{2015}).
\newblock
\showeprint[arXiv]{1503.02531}
\urldef\tempurl%
\url{http://arxiv.org/abs/1503.02531}
\showURL{%
\tempurl}


\bibitem[\protect\citeauthoryear{Lyu, Dong, Huo, and Ren}{Lyu
  et~al\mbox{.}}{2020}]%
        {DBLP:conf/aaai/LyuDHR20}
\bibfield{author}{\bibinfo{person}{Zequn Lyu}, \bibinfo{person}{Yu Dong},
  \bibinfo{person}{Chengfu Huo}, {and} \bibinfo{person}{Weijun Ren}.}
  \bibinfo{year}{2020}\natexlab{}.
\newblock \showarticletitle{Deep Match to Rank Model for Personalized
  Click-Through Rate Prediction}. In \bibinfo{booktitle}{\emph{{AAAI} 2020, New
  York, NY, USA, February 7-12, 2020}}. \bibinfo{publisher}{{AAAI} Press},
  \bibinfo{pages}{156--163}.
\newblock
\urldef\tempurl%
\url{https://aaai.org/ojs/index.php/AAAI/article/view/5346}
\showURL{%
\tempurl}


\bibitem[\protect\citeauthoryear{Ma, Wang, Zhao, Liu, Zhao, Lin, Lee, Xu, and
  Zheng}{Ma et~al\mbox{.}}{2021}]%
        {DBLP:conf/sigir/MaWZLZLLXZ21}
\bibfield{author}{\bibinfo{person}{Xu Ma}, \bibinfo{person}{Pengjie Wang},
  \bibinfo{person}{Hui Zhao}, \bibinfo{person}{Shaoguo Liu},
  \bibinfo{person}{Chuhan Zhao}, \bibinfo{person}{Wei Lin},
  \bibinfo{person}{Kuang{-}Chih Lee}, \bibinfo{person}{Jian Xu}, {and}
  \bibinfo{person}{Bo Zheng}.} \bibinfo{year}{2021}\natexlab{}.
\newblock \showarticletitle{Towards a Better Tradeoff between Effectiveness and
  Efficiency in Pre-Ranking: {A} Learnable Feature Selection based Approach}.
  In \bibinfo{booktitle}{\emph{{SIGIR} '21: The 44th International {ACM}
  {SIGIR} Conference on Research and Development in Information Retrieval,
  Virtual Event, Canada, July 11-15, 2021}}. \bibinfo{publisher}{{ACM}},
  \bibinfo{pages}{2036--2040}.
\newblock
\urldef\tempurl%
\url{https://doi.org/10.1145/3404835.3462979}
\showDOI{\tempurl}


\bibitem[\protect\citeauthoryear{McMahan, Holt, Sculley, Young, Ebner, Grady,
  Nie, Phillips, Davydov, Golovin, Chikkerur, Liu, Wattenberg, Hrafnkelsson,
  Boulos, and Kubica}{McMahan et~al\mbox{.}}{2013}]%
        {DBLP:conf/kdd/McMahanHSYEGNPDGCLWHBK13}
\bibfield{author}{\bibinfo{person}{H.~Brendan McMahan}, \bibinfo{person}{Gary
  Holt}, \bibinfo{person}{David Sculley}, \bibinfo{person}{Michael Young},
  \bibinfo{person}{Dietmar Ebner}, \bibinfo{person}{Julian Grady},
  \bibinfo{person}{Lan Nie}, \bibinfo{person}{Todd Phillips},
  \bibinfo{person}{Eugene Davydov}, \bibinfo{person}{Daniel Golovin},
  \bibinfo{person}{Sharat Chikkerur}, \bibinfo{person}{Dan Liu},
  \bibinfo{person}{Martin Wattenberg}, \bibinfo{person}{Arnar~Mar
  Hrafnkelsson}, \bibinfo{person}{Tom Boulos}, {and} \bibinfo{person}{Jeremy
  Kubica}.} \bibinfo{year}{2013}\natexlab{}.
\newblock \showarticletitle{Ad click prediction: a view from the trenches}. In
  \bibinfo{booktitle}{\emph{{KDD} 2013, Chicago, IL, USA, August 11-14, 2013}}.
  \bibinfo{publisher}{{ACM}}, \bibinfo{pages}{1222--1230}.
\newblock
\urldef\tempurl%
\url{https://doi.org/10.1145/2487575.2488200}
\showDOI{\tempurl}


\bibitem[\protect\citeauthoryear{Tian, Krishnan, and Isola}{Tian
  et~al\mbox{.}}{2020}]%
        {DBLP:conf/iclr/TianKI20}
\bibfield{author}{\bibinfo{person}{Yonglong Tian}, \bibinfo{person}{Dilip
  Krishnan}, {and} \bibinfo{person}{Phillip Isola}.}
  \bibinfo{year}{2020}\natexlab{}.
\newblock \showarticletitle{Contrastive Representation Distillation}. In
  \bibinfo{booktitle}{\emph{{ICLR} 2020, Addis Ababa, Ethiopia, April 26-30,
  2020}}. \bibinfo{publisher}{OpenReview.net}.
\newblock
\urldef\tempurl%
\url{https://openreview.net/forum?id=SkgpBJrtvS}
\showURL{%
\tempurl}


\bibitem[\protect\citeauthoryear{van~den Oord, Li, and Vinyals}{van~den Oord
  et~al\mbox{.}}{2018}]%
        {DBLP:journals/corr/abs-1807-03748}
\bibfield{author}{\bibinfo{person}{A{\"{a}}ron van~den Oord},
  \bibinfo{person}{Yazhe Li}, {and} \bibinfo{person}{Oriol Vinyals}.}
  \bibinfo{year}{2018}\natexlab{}.
\newblock \showarticletitle{Representation Learning with Contrastive Predictive
  Coding}.
\newblock \bibinfo{journal}{\emph{CoRR}}  \bibinfo{volume}{abs/1807.03748}
  (\bibinfo{year}{2018}).
\newblock
\showeprint[arXiv]{1807.03748}
\urldef\tempurl%
\url{http://arxiv.org/abs/1807.03748}
\showURL{%
\tempurl}


\bibitem[\protect\citeauthoryear{Wang, Zhao, Jiang, Zhou, Zhu, and Gai}{Wang
  et~al\mbox{.}}{2020}]%
        {DBLP:journals/corr/abs-2007-16122}
\bibfield{author}{\bibinfo{person}{Zhe Wang}, \bibinfo{person}{Liqin Zhao},
  \bibinfo{person}{Biye Jiang}, \bibinfo{person}{Guorui Zhou},
  \bibinfo{person}{Xiaoqiang Zhu}, {and} \bibinfo{person}{Kun Gai}.}
  \bibinfo{year}{2020}\natexlab{}.
\newblock \showarticletitle{{COLD:} Towards the Next Generation of Pre-Ranking
  System}.
\newblock \bibinfo{journal}{\emph{CoRR}}  \bibinfo{volume}{abs/2007.16122}
  (\bibinfo{year}{2020}).
\newblock
\showeprint[arXiv]{2007.16122}
\urldef\tempurl%
\url{https://arxiv.org/abs/2007.16122}
\showURL{%
\tempurl}


\bibitem[\protect\citeauthoryear{Xu, Li, Ge, Gao, Yang, Pei, Sun, Wu, Sun, and
  Ou}{Xu et~al\mbox{.}}{2020}]%
        {DBLP:conf/kdd/XuLGGYPSWSO20}
\bibfield{author}{\bibinfo{person}{Chen Xu}, \bibinfo{person}{Quan Li},
  \bibinfo{person}{Junfeng Ge}, \bibinfo{person}{Jinyang Gao},
  \bibinfo{person}{Xiaoyong Yang}, \bibinfo{person}{Changhua Pei},
  \bibinfo{person}{Fei Sun}, \bibinfo{person}{Jian Wu},
  \bibinfo{person}{Hanxiao Sun}, {and} \bibinfo{person}{Wenwu Ou}.}
  \bibinfo{year}{2020}\natexlab{}.
\newblock \showarticletitle{Privileged Features Distillation at Taobao
  Recommendations}. In \bibinfo{booktitle}{\emph{{KDD} '20, Virtual Event, CA,
  USA, August 23-27, 2020}}. \bibinfo{publisher}{{ACM}},
  \bibinfo{pages}{2590--2598}.
\newblock
\urldef\tempurl%
\url{https://doi.org/10.1145/3394486.3403309}
\showDOI{\tempurl}


\bibitem[\protect\citeauthoryear{Yang, Yi, Cheng, Hong, Li, Wang, Xu, and
  Chi}{Yang et~al\mbox{.}}{2020}]%
        {DBLP:conf/www/YangYCHLWXC20}
\bibfield{author}{\bibinfo{person}{Ji Yang}, \bibinfo{person}{Xinyang Yi},
  \bibinfo{person}{Derek~Zhiyuan Cheng}, \bibinfo{person}{Lichan Hong},
  \bibinfo{person}{Yang Li}, \bibinfo{person}{Simon~Xiaoming Wang},
  \bibinfo{person}{Taibai Xu}, {and} \bibinfo{person}{Ed~H. Chi}.}
  \bibinfo{year}{2020}\natexlab{}.
\newblock \showarticletitle{Mixed Negative Sampling for Learning Two-tower
  Neural Networks in Recommendations}. In \bibinfo{booktitle}{\emph{Companion
  of The 2020 Web Conference 2020, Taipei, Taiwan, April 20-24, 2020}}.
  \bibinfo{publisher}{{ACM} / {IW3C2}}, \bibinfo{pages}{441--447}.
\newblock
\urldef\tempurl%
\url{https://doi.org/10.1145/3366424.3386195}
\showDOI{\tempurl}


\bibitem[\protect\citeauthoryear{Yuan, Hsia, Yang, Zhu, Chang, Dong, and
  Lin}{Yuan et~al\mbox{.}}{2019}]%
        {DBLP:conf/cikm/YuanHYZCDL19}
\bibfield{author}{\bibinfo{person}{Bo{-}Wen Yuan}, \bibinfo{person}{Jui{-}Yang
  Hsia}, \bibinfo{person}{Meng{-}Yuan Yang}, \bibinfo{person}{Hong Zhu},
  \bibinfo{person}{Chih{-}Yao Chang}, \bibinfo{person}{Zhenhua Dong}, {and}
  \bibinfo{person}{Chih{-}Jen Lin}.} \bibinfo{year}{2019}\natexlab{}.
\newblock \showarticletitle{Improving Ad Click Prediction by Considering
  Non-displayed Events}. In \bibinfo{booktitle}{\emph{{CIKM} 2019, Beijing,
  China, November 3-7, 2019}}. \bibinfo{publisher}{{ACM}},
  \bibinfo{pages}{329--338}.
\newblock
\urldef\tempurl%
\url{https://doi.org/10.1145/3357384.3358058}
\showDOI{\tempurl}


\bibitem[\protect\citeauthoryear{Zhou, Mou, Fan, Pi, Bian, Zhou, Zhu, and
  Gai}{Zhou et~al\mbox{.}}{2019}]%
        {DBLP:conf/aaai/ZhouMFPBZZG19}
\bibfield{author}{\bibinfo{person}{Guorui Zhou}, \bibinfo{person}{Na Mou},
  \bibinfo{person}{Ying Fan}, \bibinfo{person}{Qi Pi}, \bibinfo{person}{Weijie
  Bian}, \bibinfo{person}{Chang Zhou}, \bibinfo{person}{Xiaoqiang Zhu}, {and}
  \bibinfo{person}{Kun Gai}.} \bibinfo{year}{2019}\natexlab{}.
\newblock \showarticletitle{Deep Interest Evolution Network for Click-Through
  Rate Prediction}. In \bibinfo{booktitle}{\emph{{AAAI} 2019, Honolulu, Hawaii,
  USA, January 27 - February 1, 2019}}. \bibinfo{publisher}{{AAAI} Press},
  \bibinfo{pages}{5941--5948}.
\newblock
\urldef\tempurl%
\url{https://doi.org/10.1609/aaai.v33i01.33015941}
\showDOI{\tempurl}


\bibitem[\protect\citeauthoryear{Zhou, Song, Zhu, Ma, Yan, Dai, Zhu, Jin, Li,
  and Gai}{Zhou et~al\mbox{.}}{2017}]%
        {DBLP:journals/corr/ZhouSZMYDZJLG17}
\bibfield{author}{\bibinfo{person}{Guorui Zhou}, \bibinfo{person}{Chengru
  Song}, \bibinfo{person}{Xiaoqiang Zhu}, \bibinfo{person}{Xiao Ma},
  \bibinfo{person}{Yanghui Yan}, \bibinfo{person}{Xingya Dai},
  \bibinfo{person}{Han Zhu}, \bibinfo{person}{Junqi Jin}, \bibinfo{person}{Han
  Li}, {and} \bibinfo{person}{Kun Gai}.} \bibinfo{year}{2017}\natexlab{}.
\newblock \showarticletitle{Deep Interest Network for Click-Through Rate
  Prediction}.
\newblock \bibinfo{journal}{\emph{CoRR}}  \bibinfo{volume}{abs/1706.06978}
  (\bibinfo{year}{2017}).
\newblock
\showeprint[arXiv]{1706.06978}
\urldef\tempurl%
\url{http://arxiv.org/abs/1706.06978}
\showURL{%
\tempurl}


\bibitem[\protect\citeauthoryear{Zhu, Jin, Tan, Pan, Zeng, Li, and Gai}{Zhu
  et~al\mbox{.}}{2017}]%
        {DBLP:conf/kdd/ZhuJTPZLG17}
\bibfield{author}{\bibinfo{person}{Han Zhu}, \bibinfo{person}{Junqi Jin},
  \bibinfo{person}{Chang Tan}, \bibinfo{person}{Fei Pan},
  \bibinfo{person}{Yifan Zeng}, \bibinfo{person}{Han Li}, {and}
  \bibinfo{person}{Kun Gai}.} \bibinfo{year}{2017}\natexlab{}.
\newblock \showarticletitle{Optimized Cost per Click in Taobao Display
  Advertising}. In \bibinfo{booktitle}{\emph{Proceedings of the 23rd {ACM}
  {SIGKDD} International Conference on Knowledge Discovery and Data Mining,
  Halifax, NS, Canada, August 13 - 17, 2017}}. \bibinfo{publisher}{{ACM}},
  \bibinfo{pages}{2191--2200}.
\newblock
\urldef\tempurl%
\url{https://doi.org/10.1145/3097983.3098134}
\showDOI{\tempurl}


\end{thebibliography}

%%
%% If your work has an appendix, this is the place to put it.
% \appendix

% \clearpage

% \section{Details of Evaluation Metrics}
% \textbf{G-AUC} is the weighted sum of AUC by groups. In our scenario, we calculated G-AUC as:
% \begin{equation}
%     GAUC = \frac{\sum_{(q,u)}w_{(q,u)} \times AUC_{(q,u)}}{\sum_{(q,u)}w_{(q,u)}}
% \end{equation}
% where the weights $w_{q,u}$ is proportional to the value of the query $q$ and user $u$.

% \textbf{Recall} measures the alignment of results between pre-ranking model and ranking model, which is formulated as:
% \begin{equation}
%     Recall@k = \frac{|A^{pre-ranking}_{k} \cap A^{ranking}_{k}|}{k}
% \end{equation}
% where $A^{pre-ranking}_{k}$ and $A^{ranking}_{k}$ refers to the set of top-$k$ items predicted by pre-ranking and ranking models respectively. In our experiments, we report $Recall@150$.  

% The \textbf{NDCG} metric is calculated as follows. First, we calculate the Discounted Cumulative Gain (DCG) and Ideal DCG (IDCG):
% \begin{equation}
%     DCG@k = \sum_{i=1}^{k}\frac{2^{{rel}_i}-1}{{log}_2(i+1)}, \ \ IDCG@k = \sum_{i=1}^{|{REL}_p|}\frac{2^{{rel}_i}-1}{{log}_2(i+1)}
% \end{equation}
% where $rel_i$ is the graded relevance of the result at position $i$. Then the NDCG is defined as the ratio of DCG and IDCG: $NDCG@k =DCG@k/IDCG@k$.
% % \begin{equation}
% %     NDCG@k = \frac{DCG@k}{IDCG@k}
% % \end{equation}
% In our experiments, we use $k=15$ for evaluation.

\end{document}